\documentstyle[12pt]{article}

\catcode`\@=11
\long\def\@makefntext#1{ 
\protect\noindent \hbox to 3.2pt {\hskip-.9pt
$^{{\ninerm\@thefnmark}}$\hfil}#1\hfill} 

\def\thefootnote{\fnsymbol{footnote}}
 \def\@makefnmark{\hbox to 0pt{$^{\@thefnmark}$\hss}}  

\def\ps@myheadings{\let\@mkboth\@gobbletwo
\def\@oddhead{\hbox{} 
\rightmark\hfil\ninerm\thepage}
\def\@oddfoot{}\def\@evenhead{\ninerm\thepage\hfil 
\leftmark\hbox{}}\def\@evenfoot{}
\def\sectionmark##1{}\def\subsectionmark##1{}}

\begin{document}

\newcommand{\symbolfootnote}{\renewcommand{\thefootnote}
	{\fnsymbol{footnote}}}
\renewcommand{\thefootnote}{\fnsymbol{footnote}}
\newcommand{\alphfootnote}
	{\setcounter{footnote}{0}
	 \renewcommand{\thefootnote}{\sevenrm\alph{footnote}}}

\newcounter{sectionc}\newcounter{subsectionc}\newcounter{subsubsectionc}
\renewcommand{\section}[1] {\vspace{0.6cm}\addtocounter{sectionc}{1}
\setcounter{subsectionc}{0}\setcounter{subsubsectionc}{0}\noindent
	{\bf\thesectionc. #1}\par\vspace{0.4cm}}
\renewcommand{\subsection}[1] {\vspace{0.6cm}\addtocounter{subsectionc}{1}
	\setcounter{subsubsectionc}{0}\noindent
	{\it\thesectionc.\thesubsectionc. #1}\par\vspace{0.4cm}}
\renewcommand{\subsubsection}[1]
{\vspace{0.6cm}\addtocounter{subsubsectionc}{1}
	\noindent {\rm\thesectionc.\thesubsectionc.\thesubsubsectionc.
	#1}\par\vspace{0.4cm}}
\newcommand{\nonumsection}[1] {\vspace{0.6cm}\noindent{\bf #1}
	\par\vspace{0.4cm}}

\newcounter{appendixc}
\newcounter{subappendixc}[appendixc]
\newcounter{subsubappendixc}[subappendixc]
\renewcommand{\thesubappendixc}{\Alph{appendixc}.\arabic{subappendixc}}
\renewcommand{\thesubsubappendixc}
	{\Alph{appendixc}.\arabic{subappendixc}.\arabic{subsubappendixc}}

\renewcommand{\appendix}[1] {\vspace{0.6cm}
        \refstepcounter{appendixc}
        \setcounter{figure}{0}
        \setcounter{table}{0}
        \setcounter{equation}{0}
        \renewcommand{\thefigure}{\Alph{appendixc}.\arabic{figure}}
        \renewcommand{\thetable}{\Alph{appendixc}.\arabic{table}}
        \renewcommand{\theappendixc}{\Alph{appendixc}}
        \renewcommand{\theequation}{\Alph{appendixc}.\arabic{equation}}
        \noindent{\bf Appendix \theappendixc #1}\par\vspace{0.4cm}}
\newcommand{\subappendix}[1] {\vspace{0.6cm}
        \refstepcounter{subappendixc}
        \noindent{\bf Appendix \thesubappendixc. #1}\par\vspace{0.4cm}}
\newcommand{\subsubappendix}[1] {\vspace{0.6cm}
        \refstepcounter{subsubappendixc}
        \noindent{\it Appendix \thesubsubappendixc. #1}
	\par\vspace{0.4cm}}

\def\abstracts#1{{
	\centering{\begin{minipage}{30pc}\tenrm\baselineskip=12pt\noindent
	\centerline{\tenrm ABSTRACT}\vspace{0.3cm}
	\parindent=0pt #1
	\end{minipage} }\par}}

\newcommand{\bibit}{\it}
\newcommand{\bibbf}{\bf}
\renewenvironment{thebibliography}[1]
	{\begin{list}{\arabic{enumi}.}
	{\usecounter{enumi}\setlength{\parsep}{0pt}
\setlength{\leftmargin 1.25cm}{\rightmargin 0pt}
	 \setlength{\itemsep}{0pt} \settowidth
	{\labelwidth}{#1.}\sloppy}}{\end{list}}

\topsep=0in\parsep=0in\itemsep=0in
\parindent=1.5pc

\newcounter{itemlistc}
\newcounter{romanlistc}
\newcounter{alphlistc}
\newcounter{arabiclistc}
\newenvironment{itemlist}
    	{\setcounter{itemlistc}{0}
	 \begin{list}{$\bullet$}
	{\usecounter{itemlistc}
	 \setlength{\parsep}{0pt}
	 \setlength{\itemsep}{0pt}}}{\end{list}}

\newenvironment{romanlist}
	{\setcounter{romanlistc}{0}
	 \begin{list}{$($\roman{romanlistc}$)$}
	{\usecounter{romanlistc}
	 \setlength{\parsep}{0pt}
	 \setlength{\itemsep}{0pt}}}{\end{list}}

\newenvironment{alphlist}
	{\setcounter{alphlistc}{0}
	 \begin{list}{$($\alph{alphlistc}$)$}
	{\usecounter{alphlistc}
	 \setlength{\parsep}{0pt}
	 \setlength{\itemsep}{0pt}}}{\end{list}}

\newenvironment{arabiclist}
	{\setcounter{arabiclistc}{0}
	 \begin{list}{\arabic{arabiclistc}}
	{\usecounter{arabiclistc}
	 \setlength{\parsep}{0pt}
	 \setlength{\itemsep}{0pt}}}{\end{list}}

\newcommand{\fcaption}[1]{
        \refstepcounter{figure}
        \setbox\@tempboxa = \hbox{\tenrm Fig.~\thefigure. #1}
        \ifdim \wd\@tempboxa > 6in
           {\begin{center}
        \parbox{6in}{\tenrm\baselineskip=12pt Fig.~\thefigure. #1 }
            \end{center}}
        \else
             {\begin{center}
             {\tenrm Fig.~\thefigure. #1}
              \end{center}}
        \fi}

\newcommand{\tcaption}[1]{
        \refstepcounter{table}
        \setbox\@tempboxa = \hbox{\tenrm Table~\thetable. #1}
        \ifdim \wd\@tempboxa > 6in
           {\begin{center}
        \parbox{6in}{\tenrm\baselineskip=12pt Table~\thetable. #1 }
            \end{center}}
        \else
             {\begin{center}
             {\tenrm Table~\thetable. #1}
              \end{center}}
        \fi}

\def\@citex[#1]#2{\if@filesw\immediate\write\@auxout
	{\string\citation{#2}}\fi
\def\@citea{}\@cite{\@for\@citeb:=#2\do
	{\@citea\def\@citea{,}\@ifundefined
	{b@\@citeb}{{\bf ?}\@warning
	{Citation `\@citeb' on page \thepage \space undefined}}
	{\csname b@\@citeb\endcsname}}}{#1}}

\newif\if@cghi
\def\cite{\@cghitrue\@ifnextchar [{\@tempswatrue
	\@citex}{\@tempswafalse\@citex[]}}
\def\citelow{\@cghifalse\@ifnextchar [{\@tempswatrue
	\@citex}{\@tempswafalse\@citex[]}}
\def\@cite#1#2{{$\null^{#1}$\if@tempswa\typeout
	{IJCGA warning: optional citation argument
	ignored: `#2'} \fi}}
\newcommand{\citeup}{\cite}

\def\fnm#1{$^{\mbox{\scriptsize #1}}$}
\def\fnt#1#2{\footnotetext{\kern-.3em
	{$^{\mbox{\sevenrm #1}}$}{#2}}}

\font\twelvebf=cmbx10 scaled\magstep 1
\font\twelverm=cmr10 scaled\magstep 1
\font\twelveit=cmti10 scaled\magstep 1
\font\elevenbfit=cmbxti10 scaled\magstephalf
\font\elevenbf=cmbx10 scaled\magstephalf
\font\elevenrm=cmr10 scaled\magstephalf
\font\elevenit=cmti10 scaled\magstephalf
\font\bfit=cmbxti10
\font\tenbf=cmbx10
\font\tenrm=cmr10
\font\tenit=cmti10
\font\ninebf=cmbx9
\font\ninerm=cmr9
\font\nineit=cmti9
\font\eightbf=cmbx8
\font\eightrm=cmr8
\font\eightit=cmti8


\centerline{\tenbf COLOR TRANSPARENCY EFFECTS IN QUASI-ELASTIC NUCLEAR
REACTIONS}
\baselineskip=16pt
\vspace{0.8cm}
\centerline{\tenrm GERALD A. MILLER}
\baselineskip=13pt
\centerline{\tenit Physics Department FM-15, University of Washington}
\baselineskip=12pt
\centerline{\tenit Seattle, Washington 98195, USA}
\baselineskip=13pt
\vspace{0.9cm}
\abstracts{Previous work on color transparency is reviewed briefly with an
emphasis on aspects related to  an upgrade of CEBAF.}

\vfil
\rm\baselineskip=14pt
\section{Introduction}
\vspace*{-0.7cm}

\vspace{10mm}

My CEBAF talk occurred shortly after submitting a lengthy review\cite{ARNPS}
on color transparency and the related issues of color fluctuations.
Thus the reader is directed to that review for the details.
Here I shall be concerned with presenting a brief outline and
making a few summary points. The review contains many
references, so the reference list here is short.

Usually
initial and final state
interactions cause absorptive effects which reduce
the cross sections. If color transparency CT occurs,
such interactions are suppressed at high
enough Q$^2$. I discuss some well-known examples of
possible reactions: (e,e'p), (p,pp),  (e,e'pp) and (e,e'$\Delta^{++}$).
The experimental resolution
must be good enough to insure that no extra pions are produced
and the energy transfer to  the recoil nuclear system is small
($\le 70-100 $
 MeV).
This requirement, stringent at high  energies, has hindered progress in this
field. New measurements at CEBAF and its higher energy version would be of high
interest.

CT requires the production of a point-like configuration
PLC in two-body reactions. Such PLC do not interact with
the residual nucleus.
However, even if a PLC is formed, it will expand
as it moves through the nucleus$^{2-4}$.
One can express
expansion effects in both quark and
in hadronic bases.
Such effects must be included in any realistic estimate of color transparency
effects.

The CT idea is based on: small short-lived color singlet objects are
produced in elastic hadronic reactions at high momentum transfer Q$^2$
(Sect.~2).
Such objects have small interactions with nucleons. The small system is
not an eigenstate so, unless its energy is  very high,
it expands and interacts  as it moves through the nucleus, Sect. ~3.
In addition, careful calculations including various effects present in ordinary
nuclear reactions are necessary, Sect.~4.
A summary of the implications of color transparency for the proposal of
extending CEBAF to higher energies is given in Sect.~5.

\section{Is a small system made?}  

Perhaps the most interesting question is whether or not a small
system is made in a high Q$^2$ hadronic exclusive process.
The present postulate is
that
at high Q$^2$, the
matrix elements are dominated by components or
configurations that behave as of  smaller than average
size.
Such small-sized configurations or wave packets
have been termed
point-like
configurations PLC.

If one considers asymptotically large values of Q$^2$,
perturbative QCD
holds and
PLC are produced\cite{BL80,Mue82}.
Eloquent criticisms
of  early pQCD  calculations
were put forward by Radyushkin\cite{ar}  and Isgur and Lewellyn
Smith\cite{isgurls}.

But the pQCD arguments and the criticisms thereof were
not complete because one must
investigate the possible role of
low momentum (soft) long wavelength
gluons that are radiated as the colored quarks are
accelerated. The
effects of such radiation can be included
via a form factor similar to that introduced by Sudakov,
which
 decreases the probability for elastic scattering of a free fermion.
However,
for a color singlet system,
 the gluon radiation contributions
cancel if the quarks and gluons making up the system are closely separated.
Then significant
contributions to the elastic form factor
occur mainly for configurations of small size.
The so-called Sudakov effects were known early on
but
numerical evaluations
  did not occur
until recently with the work of Botts, Li and
Sterman$^{10-13}$ and now others.

So far we have discussed pQCD calculations. But if one is interested in
seeing how color transparency effects grow as Q$^2$ is increased from
low values it is necessary to see if
non-perturbative calculations also admit a PLC.
Several different models have been examined using a new
numerical criterion\cite{FMS92,FMS931}. The result
is that the form factor is dominated
by PLC within
many non-perturbative models. Furthermore, these effects set in at relatively
low values of
 Q$^2$.

\section{Time development}

Suppose a $PLC$ is produced in the interior of the nucleus. Any
non-eigenstate
undergoes time development. Here expansion occurs because the
starting system is defined to be small.
This
expansion has been found to be a vital effect
for intermediate energies, $P_{lab}$
less than about 20 GeV/c.

We are concerned here  with time development in
nuclear quasielastic reactions. Consider
the (e,e'p) reaction.
The virtual photon is absorbed by a proton
creating a high momentum object which is ejected from the nucleus.
The old fashioned approach is to treat the ejectile as proton. Then the
final state interactions are governed by the optical potential $U^{opt}$.
If the
proton wave function is computed from
$U^{opt}$, the proton wave is said to be
distorted (from the plane wave approximation). The use of
such a wave function in computing cross sections is called
the distorted wave impulse approximation DWIA, where the ``impulse"
refers to the use of the free nucleon-nucleon cross section.

But if the ejected object is a PLC, using
U$^{opt}$ is not appropriate. On the other hand, the ejectile expands as it
moves through the nucleus, so that one can not simply
neglect the soft final state interactions.  The need to include
this expansion
was recognized by Farrar et al.\cite{FLFS89} who
argued that the square of the transverse size (and therefore the
forward scattering amplitude) is roughly
proportional to  the
distance travelled
$Z$ from the point of hard interaction where the PLC is formed.

The time development of the $PLC$ can also be
obtained by modeling the ejectile-nucleus interaction as
$\hat U = - i \sigma (b^2) \rho (R),$ where $b^2$ represents the
transverse separation of the quarks and
$\rho(R)$ is the nuclear density at
a distance R from the nuclear center.
Then
one can assume a baryonic basis and compute the relevant matrix elements
of $\sigma (b^2)$.
Jennings and Miller\cite{jm90,jm91}
solved the Lippman-Schwinger equation
using an
exponentiating procedure.
Greenberg and
Miller\cite{greenberg} showed that exponentiation
is often a very  accurate approximation.
A more elaborate approach was taken later\cite{jm92} by using
measured matrix elements for deep inelastic scattering
and diffractive
dissociation.
In this case, an approximate linear growth of the PLC cross section with
distance is obtained.

Still another approach involves
treating the baryon-nucleon amplitude in terms of a
finite number of baryonic states.
Then the baryon-nucleon T-matrix can be represented as an
N by N matrix.
The eigenvalues of such a matrix are an example of the Good-Walker diffractive
eigenvalues\cite{GW60}.
The absence of interactions required for complete
color transparency can only be
obtained if the T-matrix has at least one state of eigenvalue
0, the PLC\cite{FGMS92},
but several different papers
use two state models without satisfying this
condition.

\section{Relevant data}
\label{subs:search}
It is natural to consider the (e,e'p) and (p,pp) processes for color
transparency searches.
The first published experiment aimed at
color transparency was the BNL $(p,pp)$
work of Carroll {\it et al.}\cite{bnl88}. The only other one published
is the SLAC (e,e'p) NE18 experiment\cite{ne18}.
We discuss each.

\subsection{The BNL (p,pp) experiment}
\vspace*{-0.35cm}
Proton beams of
momenta $p_L$ were 6, 10 and  12 GeV/c  aimed at a target consisting of
CH$_2$ interleaved with nuclei.
The experimental setup used was that for  proton hydrogen
 elastic scattering
at a center of mass angle of
90$^\circ$.
For the hydrogen target, identifying an elastic scattering event
requires  detecting  the outgoing
momentum of one proton and the angle of the other.
This is not sufficient for nuclear scattering because of the
Fermi motion of the bound proton.
However, information from veto counters was used to suppress
inelastic events.

The data were originally plotted with an effective beam momentum,
$P_{eff}$. It is better to use
-$k_z$, the component of the momentum
of the struck nucleon calculated in the plane wave impulse
approximation.  The z-axis is defined by the beam direction.
The DWIA describes similar data at intermediate energies of $E_p= $1 GeV
with accuracy of better than $20\%$.
However the BNL data
are considerably above the DWIA
results, see
Figs.~9  and 10 of Ref.~[1].

The observed large
value of the ratio of the cross section to its value in Born approximation
indicated
the presence of a large transparency effect, but the
apparent drop at 12 GeV/c caused considerable discussion.\cite{bdet,rp88}
The color transparency
models which include expansion effects naturally produce an increase  of
the transparency consistent with the one observed in  the BNL experiment
at 6 and 10  GeV/c.

The possible
nuclear results
depend on the
pp elastic scattering data. The energy dependence of the
90$^\circ$ angular distribution is of the form of 1/s$^{10}$ R(s)
where R(s) oscillates between 1 and 3 over the energy range of the BNL
experiment.
Ralston and Pire\cite{rp88}, suggested that the energy
dependence is caused by an interference between an amplitude which produces
a PLC, and a soft one
which involves a
large or blob-like
configuration BLC.
Another mechanism  
is that of Brodsky and de Teramond.\cite{bdet}
It is natural to discuss high Q$^2$ elastic proton-proton scattering
in terms of configurations of different sizes.
Separating the contributing configurations into two, a PLC and a BLC
is only a simple first step.

Effects of the Fermi motion in treating the expansion process were
evaluated\cite{boffi,kj93,gardner}, and big numerical effects
were obtained.
The result was
that it became possible to construct a model\cite{jm93} which is
 able to describe BNL data at all energies.
The
solid curves of Fig.~10 of Ref.~[1]
show the full calculation of Ref.~[26]
including the Ralston-Pire
intereference effect.
Keeping the
experimental uncertainties in mind, the agreement between theory and
experiment is rather good.

\subsection{The SLAC (e,e'p) experiment}
\label{subs:slac}
\vspace*{-0.35cm}
If color transparency effects observed at BNL
are real they should be manifest in reactions
other  than (p,pp).
Thus the
recent measurement of the (e,e'p) reaction
 made at SLAC\cite{ne18} and the possibility of future work at CEBAF
are very exciting.
 The NE18
collaboration measured cross sections for $^{12}$C, Fe and Ag targets
for momentum transfers Q$^2$ of 1, 3, 5 and 6.8 GeV$^2$. We quote results
presented recently\cite{ne18}
for $^{12}$C.
The available data and the related theory are shown in Fig.~11
of Ref.~[1].

The NE-18 experiment has made a significant achievement in
observing the
quasi-elastic (e,ep') reaction at Q$^2$ between 1 and 7 GeV$^2$.
 One is now faced with the
task of assessing
the data.
The results of Ref.~[20]
are that no significant rise of the transparency with Q$^2$ is seen.
Early predictions depended on the unknown expansion rate.  This is
still unknown but is now better constrained by the (p,pp) data.
One is now in a much better position to ask for the SLAC experiment:
how large can one expect CT effects to be?
One way to see is to
compare the data with DWIA calculations, another way is to
use models consistent with the BNL data
to compute CT effects for the (e,e'p) reaction.

Relevant DWIA calculations
must satisfy certain criteria:
(1) compute the relevant observable T(A) according to the experimental
acceptance;
(2) use nuclear wave functions which
reproduce the nuclear density
and spectral function;
(3) include the energy dependence  of $\sigma$.
Calculations satisfying  these criteria are shown in Fig.~11 of
Ref.~[1] along with calculations including color transparency effects
consistent with the BNL results.

The net result is that calculations which predict substantial color
transparency
effects for the (p,pp) reaction do not predict much color transparency
in the regime of  Q$^2$ available to the NE18 experiment.
The (e,e'p) reaction is inherently simpler than the (p,pp) reaction
so it is imperative to push the (e,e'p) measurements to higher values
of Q$^2$, say up to 12 or 15 GeV$^2$.  Observing substantial effects would be
possible.$^{16}$

\subsection{Rescattering vs. time development}
\vspace*{-0.35cm}
The problem in looking for CT effects in experiments at $Q^2$
from about one to a few GeV$^2$ is that the assumed $PLC$ expands rapidly
while propagating through the nucleus. To observe CT at
intermediate values of $Q^2$ it is necessary to suppress this
expansion.
If one studies a process where the produced
system can {\bf only} be produced
by an interaction  in the final state, a
double scattering event, then the color coherent effects would be manifest
as a decrease of the probability for final-state interactions
with increasing Q$^2$.  One could then observe an effect decreasing from
the value expected without CT (Glauber-value) to zero.  Thus, the measured
cross section is to be compared with a vanishing quantity so that the
relevant ratio of cross sections runs from 1 to infinity.
The first calculations\cite{double} show that substantial CT effects
are observable in the
(e,e'pp) reactions on $^{4,3}He$ targets. Such experiments can be done at
CEBAF.
It would be important to do these experiments at low Q$^2$, say 4 to 6 GeV$^2$
to establish the effects and then to confirm them by going to higher values of
Q$^2$.

Another idea involves pionic degrees of freedom.
Probing a nucleon at intermediate momentum transfers ($Q^2$ about a
few GeV$^2$) may produce a small system without a
pion cloud\cite{FMS92,FMS931}.
This
cloud-stripping effect can be studied
by considering processes that require a pion
exchange to proceed. An example is the quasielastic production of the
$\Delta^{++}$ in electron scattering - the $(e,e^\prime\Delta^{++})$
reaction. The initial singly charged object is knocked out of the nucleus
by the virtual photon and converts to a $\Delta^{++}$ by emitting or
absorbing a charged pion. But pionic coupling to small-sized systems is
suppressed, so  this cross section for quasielastic production of
$\Delta^{++}$'s should fall faster with increasing  $Q^2$ than the
predictions of conventional theories. Calculations are now in progress
\cite{FLMS94}.
This could be a new  kind of transparency that
involves pions, so
 the name ``chiral transparency"\cite{FMS92} was invoked.

\section{Spin Dependent Color Transparency}

Bill Greenberg's thesis included a detailed study of effects
which can be observed by measuring the polarization of the outgoing proton.
I will discuss the principle results of this work\cite{GM94}.

Our procedure was to treat the
vector nature of the photon and the
spin of the proton and photon
explicitly by describing the initial bound proton and the ejected wave
packet as four-component Dirac spinors.
Such effects, ignored in previous calculations, yield several results.

(1)
The use of Dirac based optical potentials in
the ``standard calculation" (ignoring CT)  leads to smaller
cross sections than predicted before.  This model dependence,
which arises from the different radial form used in the optical potential,
means that one must be
wary of claims that color transparency can be observed by finding small
differences between data and the magnitude of a DWIA prediction.

(2)
The normal component of the
ejectile polarization, which vanishes in the limit of full CT, is found to
approach zero very slowly as the energy increases. Finding color transparency
by measuring the polarization of the outgoing proton will be very difficult at
{\em any} forseeable energy.

(3)
The presence of the $1H_{11/2}$ orbital, causes
the normal-transverse response in
$^{208}Pb$ to be
sensitive to color transparency effects
at quite low
momentum transfers $\sim$ 1~GeV/c.  There is a similar effect in $^{120}$Sn.

(4)
The four-component nature of our formalism, allows us to determine
that our calculations are roughly consistent with current conservation,
except when the momentum $k_z$
of the struck nucleon is greater than about 150~MeV/c. Here the $z$ direction
is that of the virtual photon.
More generally, we argue that
attempts to enhance color transparency
effects by measuring cross sections  for large values of $k_z$
are risky. One needs to check that the relevant predictions satisfy the current
conservation.

\section{Implications for CEBAF at higher energy}

Studying the (e,e'p) reaction at high Q$^2$ represents an excellent opportunity
to observe color transparency.
One should not be discouraged by the NE18 results.
Calculations which predict substantial color transparency
effects for the (p,pp) reaction do not predict much color transparency
in the regime of  Q$^2$ available to the NE18 experiment.
The (e,e'p) reaction is inherently simpler than the (p,pp) reaction
so it is imperative to push the (e,e'p) measurments to higher values
of Q$^2$, say up to 12 or 15 GeV$^2$.
If the effect is not seen at 15 GeV$^2$, it can regarded as inconsequential.

Using double scattering reactions (e,e'pp) and (e,e'$\Delta^{++}$)
would allow a measurement of color transparency effects at modest values of
Q$^2$, from 1 or 2 up to about 6 GeV$^2$.
It would be important to do these experiments
to establish the effects and then to confirm them by going to higher values of
Q$^2$.

Observing color transparency in (e,e'$\vec p$) measurements would be very
 difficult at any momentum transfer. Such measurements are sensitive to
interesting nuclear structure effects, but there is no urgent need to do these
at higher energies.

\section{Acknowledgments}
The work discussed in my oral
CEBAF presentation  was done in collaboration with
 K.S. Egiyan, L. Frankfurt, W.R.~Greenberg, B.K.  Jennings, M.M. Sargsyan,
and M.Strikman.

This work was supported in part by the USDOE.


\section{References}
\begin{thebibliography}{99}

\bibitem{ARNPS} L.L. Frankfurt, G.A. Miller and M. Strikman,
``The Geometrical Optics of Coherent High Energy Processes",
U. Wa preprint DOE/ER/40427-06-N94; in press
{\it Ann. Rev. Nucl. Part. Sci.} {\bf 44} (1994).

\bibitem {FLFS89} G.R.~Farrar,
H.~Liu, L.L.~Frankfurt \& M.I.~Strikman, {\it Phys.
Rev. Lett.} {\bf  61} (1988) 686.

\bibitem{jm90}
B.K.~Jennings and G.A.~Miller, {\it Phys. Lett.} {\bf B236} (1990) 209.

\bibitem{jm91}
  B.K.~Jennings and G.A.~Miller, {\it Phys. Rev.} {\bf D44} (1991)
692.

\bibitem{greenberg} W.R.~Greenberg and G.A.~Miller, {\it Phys. Rev.}
{\bf D47}
(1993) 1865.

\bibitem{BL80} S.J. Brodsky, G.P. Lepage, {\it Phys. Rev.} {\bf D} (1980) 2157.

\bibitem {Mue82} A.H.~Mueller in {\it Proceedings of Seventeenth rencontre de
    Moriond,  1982} ed. J Tran Thanh Van (Editions Frontieres,
Gif-sur-Yvette, France, 1982) Vol. I, p13.

\bibitem{ar}
 G.P. Korchemski, A.V. Radyushkin,
{\it Sov. J. Nucl. Phys.} {\bf 45} (1987) 910 and refs. therein.

\bibitem {isgurls} N.~Isgur and C.H.~Llewellyn-Smith, {\it Phys. Rev. Lett.}
{\bf  52} (1984)
1080
{\it Phys. Lett.} {\bf B217} (1989) 535.

\bibitem{listerman}J. Botts and G. Sterman,
{\it Nucl. Phys.} {\bf B325} (1989) 62.

\bibitem{Li}
 H-N
Li {\it Phys. Rev.} {\bf  D48} (1993) 4243.

\bibitem{Botts}J. Botts and G. Sterman, {\it Nucl. Phys.} {\bf B325}, 62
(1989).

\bibitem{sterman}H.-N. Li and G. Sterman, {\it Nucl. Phys.} {\bf B381}
(1992) 129.

\bibitem {FMS92}
L. Frankfurt, G.A. Miller, and M. Strikman, {\it Comments on Nuclear
and Particle Physics} {\bf 21} (1992) 1.

\bibitem{FMS931}   L. Frankfurt, G.A. Miller, and  M.Strikman,
{\it Nucl. Phys.} {\bf A555} (1993) 752.

\bibitem{jm92} B.K.~Jennings and G.A.~Miller,
{\it Phys. Rev. Lett.} {\bf 70} (1992) 3619; {\it Phys. Lett.} {\bf B274}
(1992) 442.

\bibitem{GW60} M.L. Good and W.D.  Walker, {\it Phys.  Rev.}  {\bf 120},
(1960) 1857.

\bibitem{FGMS92}
L.~Frankfurt,
W.~R.~Greenberg,
G.~A.~Miller and~M.~Strikman, {\it Phys. Rev.} {\bf C46} (1992) 2547.

\bibitem{bnl88} A.S.~Carroll et al. {\it Phys. Rev. Lett.} {\bf 61} (1988)
1698.

\bibitem {ne18}  N.C.R. Makins et. al.
{\it Phys. Rev. Lett.} {\bf 72} (1994)  1986.

\bibitem {bdet}S.J.~Brodsky \& G.F.~De~Teramond, {\it Phys. Rev. Lett.}
{\bf  60}
(1988) 1924.

\bibitem {rp88} J.P.~Ralston and B.~Pire, {\it Phys. Rev. Lett.} {\bf 61}
 (1988) 1823.

\bibitem{boffi} S. Boffi, D.E. Kharzeev
{\it Phys. Lett.} {\bf B305} (1993) 1;
{\it Nucl.Phys.} {\bf A565} (1993) 767;
{\bf B325} (1994) 294.

\bibitem{kj93} B.K.
 Jennings and B. Kopeliovich, {\it Phys. Rev. Lett.} {\bf 70} (1993) 3384.

\bibitem{gardner}
 S. Gardner
``Color transparency in (e,e'p) and the electroproduction of resonances"
IU-NTC-92-27, Aug 1992.

\bibitem{jm93}B.K.~Jennings and G.A. Miller, {\it Phys. Lett.} {\bf B318}
 (1993) 7.

\bibitem{double}  K.S. Egiyan,  L.L.
Frankfurt,  W.R. Greenberg,  G.A. Miller,   M.M. Sargsyan,
     M.I. Strikman, Searching for
Color Coherence Effects Via Double Scattering Events,
1993 preprint, in press {\it Nucl. Phys} {\bf A}.

\bibitem{FLMS94} L. Frankfurt, T.-S.H. Lee, G. A. Miller and M.Strikman,
in progress.

\bibitem{GM94} W.R. Greenberg and G.A.  Miller, {\it Phys. Rev.}
{\bf C49} (1994) 2747; Erratum Phys. Rev. C;
W.R.~Greenberg, 1993 U. Wa. Ph. D. thesis. ``Color Transparency
in Quasi-Elastic (e,e'p) Reactions at Large Momentum Transfers"

\end{thebibliography}
 \end{document}